\documentstyle[12pt,bezier]{article}
\begin{document}
\newcommand{\de}{\delta}\newcommand{\ga}{\gamma}
\newcommand{\e}{\epsilon} \newcommand{\th}{\theta}\newcommand{\ot}{\otimes}
\newcommand{\ba}{\begin{array}} \newcommand{\ea}{\end{array}}
\newcommand{\beq}{\begin{equation}}\newcommand{\eeq}{\end{equation}}
\newcommand{\tmod}{{\cal T}}\newcommand{\amod}{{\cal A}}
\newcommand{\bemod}{{\cal B}}\newcommand{\cmod}{{\cal C}}
\newcommand{\dmod}{{\cal D}}\newcommand{\hmod}{{\cal H}}
\newcommand{\s}{\scriptstyle}\newcommand{\tr}{{\rm tr}}
\newcommand{\einsop}{{\bf 1}}
\title{Quantum Group Invariant Supersymmetric t-J Model
with Periodic Boundary Conditions}
\author{Angela Foerster\thanks{e-mail: angela@if.ufrgs.br}}
\date{\small\it Instituto de F\'{\i}sica da Universidade Federal do
Rio Grande do Sul\\ Av. Bento Gon\c{c}alves 9500, 91.501-970 Porto
Alegre, RS \\ Brazil}
\maketitle
\begin{abstract}
An integrable version of the supersymmetric t-J model which is
quantum group invariant as well as periodic is introduced and
analysed in detail. The model is solved through the algebraic
nested Bethe ansatz method. \end{abstract}

\newpage
The Bethe Ansatz Method  \cite{at}, first introduced to solve the
$XXX$ Heisenberg chain, is one of the most powerful tools in the
treatment of integrable models. Its further development had
important contributions from Yang and Yang \cite{bt} and Baxter
\cite{ct}, among others (for a review, see De Vega \cite{dt}). A
great impulse in the theory of integrable systems was given by the
Quantum Inverse Scattering Method  \cite{et}. This approach
provides a unified framework for the exact solutions of classical
and quantum models and led naturally to the new mathematical
concept of quantum groups \cite{ft}.
The construction of quantum group invariant integrable models has
been attracting considerable attention. One possibility to obtain such
invariant models is to deal with open boundary conditions (OBC). In
this connection, some quantum group invariant integrable models,
such as the XXZ-Heisenberg model \cite{Nep1}, \cite{Des}, the
$spl_q(2,1)$ supersymmetric t-J model \cite{For2}, \cite{Gr1}, the
$SU_q(N)$ \cite{Gr2}, the $SU_q(n/m)$ \cite{MR}, \cite{Rui} and the
$B_n^{(1)}, C_n^{(1)} $ and $D_n^{(1)}$ spin chains \cite{Nep2}
have been formulated. In particular, ( except for the
$B_n^{(1)},C_n^{(1)} $ and $D_n^{(1)}$ cases ) their spectrum
have been obtained through a
generalization of the Sklyanin-Cherednik construction of the Yang-
Baxter algebra \cite{Cher}, \cite{skly}. For these cases, the
use of OBC turned
the calculations much more complex than for periodic boundary
conditions (PBC). For instance, the commutation relations between
the pseudoparticle operators ${\cal B}_{\alpha}$ and the transfer
matrix are much involved. In addition, the structure of the
unwanted terms generated in the procedure is
so complicated that just after sophisticated manipulations it is
possible to recognize wanted and unwanted contributions. For the
$B_n^{(1)},C_n^{(1)} $ and $D_n^{(1)}$ chains, due to
technical difficulties, a "doubled" postulate has been proposed to
obtain the spectrum.

Recently, the question whether the quantum group invariance
necessarily implies the use of OBC has been addressed in the
literature. The construction of quantum group invariant integrable
closed chains has been examined by some authors \cite{Mar},
\cite{Zap}, \cite{Grosse} and, in fact, a
quantum group invariant XXZ model and
an $U_q(sl(n))$ invariant chain with PBC have been
formulated and analysed in detail.
Therefore, it is of interest to find other quantum group invariant
integrable closed chains.

In this paper we introduce an integrable version of the
supersymmetric t-J model which is quantum group invariant and
periodic. The system is of interest because of its possible
connection with high-Tc superconductivity. It describes electrons
with nearest-neighbor hopping and spin exchange interaction on a chain
( see eq. (\ref{18}) ) and can be considered as an anisotropic extension
of the supersymmetric t-J model. Its physical properties are, of
course, essentially the same as for the case of OBC. Nevertheless,
the approach here adopted simplifies drastically the nested Bethe ansatz.
Moreover, this is the first time that a quantum supergroup invariant
integrable periodic model is presented.
The corresponding Hamiltonian is related to a transfer matrix of a
"graded" vertex model \cite{KS} with anisotropy.
Through a generalization
of the construction of ref.\cite{Zap} to the case of a "graded"
15-vertex model the system is analysed and the Bethe
ansatz equations are obtained.

We begin by introducing the $R$-matrix, which in terms of a generic
spectral parameter $x$ and a deformation parameter $q$ reads
\cite{Zhan}
\begin{equation}
\label{1}
R^{\gamma \delta}_{\alpha \beta}(x)=
\ba{c}
\unitlength=0.50mm
\begin{picture}(20.,25.)
\put(-3.,2.){\makebox(0.,0.){$\s x$}}
\put(0.,-4.){\makebox(0.,0.){$\s \beta $}}
\put(23.,2.){\makebox(0.,0.){$\s 1$}}
\put(20.,-4.){\makebox(0.,0.){$\s \alpha $}}
\put(0.,24.){\makebox(0.,0.){$\s \gamma $}}
\put(20.,24.){\makebox(0.,0.){$\s \delta $}}
\put(20.,0){\vector(-1,1){20.}}
\put(0.,0){\vector(1,1){20.}}
\put(0.,0.){\line(1,1){20.}}
\put(0.,20.){\line(1,-1){20.}}
\end{picture}
\ea =
\pmatrix{a&0&0&|&  0&0&0&|&  0&0&0\cr
               0&b&0&|&  c_-&0&0&|&  0&0&0\cr
               0&0&b&|&  0&0&0&|&  c_-&0&0\cr
               -&-&-& &  -&-&-& &  -&-&-\cr
               0&c_+&0&|&  b&0&0&|&  0&0&0\cr
               0&0&0&|&  0&a&0&|&  0&0&0\cr
               0&0&0&|&  0&0&b&|&  0&c_-&0\cr
               -&-&-& &  -&-&-& &  -&-&-\cr
               0&0&c_+&|&  0&0&0&|&  b&0&0\cr
               0&0&0&|&  0&0&c_+&|&  0&b&0\cr
               0&0&0&|&  0&0&0&|&  0&0&w\cr} \, \, \, ,
\end{equation}
where $\alpha$ , $\beta$ ( $\gamma$ and $\delta$ ) are column ( row
) indices running from 1 to 3 and
\begin{equation}
\label{2}
a = x q - { 1 \over x q } ,\quad
b = x - { 1 \over x } ,\quad
c_+ = x ( q - { 1 \over q} ), \quad
c_- = { 1 \over x } (q - { 1 \over q} ),\quad
w = -{ x \over q}  + { q \over x }\, \, \, .
\end{equation}
The subscript $x$ and $1$ in the figure will soon become clear. The
$R$-matrix~(\ref{1}) acts in the tensor product of two
3-dimensional auxiliary spaces ${\bf C}^3 \otimes {\bf C}^3$ and it
fulfills the Yang-Baxter equation
\begin{equation}
\label{3}
R^{\alpha^{\prime\prime} \beta^{\prime\prime} }_{\alpha^\prime
\beta^\prime}(x/y)
R^{\alpha^\prime \gamma^{\prime\prime} }_{ \alpha \phantom{0}
\gamma^\prime }(x)
R^{\beta^\prime \gamma^\prime}_{\beta \phantom{0} \gamma}(y) =
R^{\beta^{\prime\prime} \gamma^{\prime\prime} }_{\beta^\prime
\gamma^\prime}(y)
R^{\alpha^{\prime\prime} \gamma^\prime}_{\alpha^\prime \phantom{0}
\gamma}(x)
R^{\alpha^\prime \beta^\prime}_{\alpha \phantom{0} \beta}(x/y) \,
\, . \end{equation}
It is easy to check that it also satisfies the Cherdnik's
reflection property \cite{Cher}
\begin{equation}
\label{4}
R^{\alpha \phantom{0}  \beta \phantom{0}
}_{\alpha^\prime\beta^\prime  }(x/y)
R^{\beta^\prime\alpha^\prime}_{\gamma \phantom{0}  \delta
\phantom{0}} ({\mu \over x y } )=
R^{\alpha\phantom{0}  \beta \phantom{0}  }_{\alpha^\prime 
\beta^\prime }({ \mu \over x y} ) R^{\beta^\prime  \alpha^\prime 
}_{\gamma \phantom{0} \delta \phantom{0}  }(x/y) \, \, \, \, \,
\ba{c}
\unitlength=0.50mm
\begin{picture}(100.,70.)
\put(20.,10.){\line(-1,2){10.}}
\put(10.,45.){\line(3,-1){30.}}
\put(92.,13.){\makebox(0,0)[cc]{$\s x$}}
\put(85.,65.){\makebox(0,0)[cc]{$\s \mu /x$}}
\put(34.,32.){\makebox(0,0)[cc]{$\s y$}}
\put(100.,40.){\makebox(0,0)[cc]{$\s \mu /y$}}
\put(55.,40.){\makebox(0,0)[cc]{$=$}}
\put(70.,50.){\line(1,-2){20.}}
\put(70.,35.){\line(3,-1){30.}}
\put(94.,22.){\makebox(0,0)[cc]{$\s y$}}
\put(22.,14.){\makebox(0,0)[cc]{$\s x$}}
\put(34.,54.){\makebox(0,0)[lc]{$\s \mu /y$}}
\put(40.,67.){\makebox(0,0)[lc]{$\s \mu /x$}}
\put(10.,30.){\vector(3,4){30.}}
\put(10.,45.){\vector(2,1){30.}}
\put(70.,35.){\vector(3,1){30.}}
\put(90.,7.){\makebox(0,0)[cc]{$\s \delta$}}
\put(20.,6.){\makebox(0,0)[cc]{$\s \delta$}}
\put(42.,33.){\makebox(0,0)[cc]{$\s \gamma$}}
\put(102.,23.){\makebox(0,0)[cc]{$\s \gamma$}}
\put(82.,73.){\makebox(0,0)[cc]{$\s \alpha$}}
\put(42.,72.){\makebox(0,0)[cc]{$\s \alpha$}}
\put(103.,46.){\makebox(0,0)[cc]{$\s \beta$}}
\put(43.,60.){\makebox(0,0)[cc]{$\s \beta$}}
\put(10.00,30.){\circle{1.5}}
\put(11.00,45.){\circle{1.5}}
\put(70.00,50.){\circle{1.5}}
\put(70.00,35.){\circle{1.5}}
\put(70.,50.){\vector(1,2){10.}}
\end{picture}
\ea.
\end{equation}
Here the symbol ($\circ$) indicates that at this point the
spectral parameter changes from $x$ to $\mu/x$ and $y$ to $\mu/y$.
Notice the presence of an arbitrary constant $ \mu $ in the above
equation, which is related with the choice of the boundaries. As in
the case of the $U_q(sl(n))$ invariant integrable
chain \cite{Zap}, we will take the limit $\mu\to\infty$
in order to construct a quantum group invariant model with
PBC.

For later convenience the spectral parameter dependent $R$-
matrix~(\ref{1}) can be written in terms of "constant" $R$-
matrices ( $R_{\pm} $) as
\begin{equation}
\label{5}
R(x) \, = \, x \, R_+ \, - \, \frac{1}{x} \, R_- \, \, = \, \, x \,
\ba{c} \unitlength=0.3mm
\begin{picture}(30.,30.)
\put(30.,0.){\vector(-1,1){30.}}
\put(18.,18.){\vector(1,1){12.}}
\put(0.,0.){\line(1,1){12.}}
\end{picture}
\ea
\, \, - \, \, \frac{1}{x} \,  \ba{c}
\unitlength=0.3mm
\begin{picture}(30.,30.)
\put(0.,0.){\vector(1,1){30.}}
\put(12.,18.){\vector(-1,1){12.}}
\put(30.,0.){\line(-1,1){12.}}
\end{picture}
\ea\ ,
\end{equation}
where $R_+$ ($R_-$) corresponds to the leading term
in the limit of the matrix $R(x)$ for $x
\rightarrow \infty  (0)$.

As usual, the standard monodromy matrix is defined as the product
of $R$-matrices~(\ref{1}) as follows
\begin{equation}
\label{6}
T_{\alpha \{ \beta \} }^{\gamma \{ \delta \} }(x)=
R^{\gamma \,\, \delta_1}_{\alpha_1 \beta_1}(1/x)
R^{\alpha_1 \delta_2}_{\alpha_2 \beta_2}(1/x)
\ldots R^{\alpha_{L-1} \delta_L}_{\alpha \,\, \beta_L}(1/x)= \ba{c}
\unitlength=0.50mm
\begin{picture}(99.,29.)
\put(64.,15.){\line(1,0){24.}}
\put(55.,15.){\makebox(0,0)[cc]{$\cdots$}}
\put(32.,-5.){\makebox(0,0)[cc]{$\s \beta_1$}}
\put(42.,-5.){\makebox(0,0)[cc]{$\s \beta_2$}}
\put(77.,-5.){\makebox(0,0)[cc]{$\s \beta_L$}}
\put(32.,5.){\makebox(0,0)[cc]{$\s 1$}}
\put(43.,5.){\makebox(0,0)[cc]{$\s 1$}}
\put(78.,5.){\makebox(0,0)[cc]{$\s 1$}}
\put(32.,32.){\makebox(0,0)[cc]{$\s \delta_1$}}
\put(42.,32.){\makebox(0,0)[cc]{$\s \delta_2$}}
\put(77.,32.){\makebox(0,0)[cc]{$\s \delta_L$}}
\put(47.,15.){\vector(-1,0){31.}}
\put(92.,15.){\makebox(0,0)[cc]{$\s \alpha$}}
\put(85.,17.){\makebox(0,0)[cc]{$\s x$}}
\put(12.,15.){\makebox(0,0)[rc]{$\s \gamma$}}
\put(30.,0.){\vector(0,1){28.}}
\put(40.,0.){\vector(0,1){28.}}
\put(75.,0.){\vector(0,1){28.}}
\end{picture}
\ea  .
\end{equation}
It acts in the tensor product of a
$L$-dimensional "quantum space" and a 3-dimensional
auxiliary space ( $ {\bf C}^{3L} \times {\bf C}^{3} $) .
For the
case $ q= 1 $, taking the trace of the
$ T-$ matrix~(\ref{6}) in the auxiliary space one gets
an $spl(2,1)$-invariant transfer matrix, related with
the supersymmetric t-J model \cite{For1}. However, for
$ q \neq 1 $, this trace does not generate an
$ spl_q(2,1)$-invariant transfer matrix. Then,
in order to construct a quantum group invariant integrable model we
have to introduce the "doubled" monodromy matrix $\cal{U}$
\begin{equation}
\label{7}
{\cal{U}}^{\gamma \{ \delta \} }_{\alpha \{ \beta \} }
(x, \{ \mu \} ) \, \, = \, \, {\tilde{T}}^{\gamma \{
\delta \} }_{\alpha^\prime \{
\beta^\prime \}}(\mu/x) \, {T}^{\alpha^\prime \{ \beta^\prime \}
}_{\alpha \{ \beta\} }(x)
\, = \, \, \,
\ba{c}
\unitlength=0.50mm
\begin{picture}(95.,49.)
\put(45.,15.){\makebox(0,0)[cc]{$\cdots$}}
\put(21.,-3.){\makebox(0,0)[cc]{$\s \beta_1$}}
\put(31.,-3.){\makebox(0,0)[cc]{$\s \beta_2$}}
\put(71.,-3.){\makebox(0,0)[cc]{$\s \beta_L$}}
\put(21.,47.){\makebox(0,0)[cc]{$\s \delta_1$}}
\put(31.,47.){\makebox(0,0)[cc]{$\s \delta_2$}}
\put(71.,47.){\makebox(0,0)[cc]{$\s \delta_L$}}
\put(86.,15.){\makebox(0,0)[cc]{$\s \alpha$}}
\put(15.,25.){\oval(20.,20.)[l]}
\put(70.,0.){\vector(0,1){45.}}
\put(30.,0.){\vector(0,1){45.}}
\put(20.,0.){\vector(0,1){45.}}
\put(73.,35.){\vector(1,0){10.}}
\put(57.,35.){\line(1,0){25.}}
\put(57.,15.){\line(1,0){25.}}
\put(45.,35.){\makebox(0,0)[cc]{$\cdots$}}
\put(86.,35.){\makebox(0,0)[cc]{$\s \gamma$}}
\put(7.,34.){\makebox(0,0)[rc]{$\s \mu/x$}}
\put(4.,17.){\makebox(0,0)[rc]{$\s x$}}
\put(15.,15.){\line(1,0){21.}}
\put(15.,35.){\line(1,0){21.}}
\put(5.00,25.){\circle{1.5}}
\end{picture}
\ea\ ,
\end{equation}
where $\tilde{T}$ is a row-to-row monodromy matrix proportional to
$T^{-1}$
\begin{equation}
\label{8}
\tilde{T}_{\alpha \{ \beta \} } ^{ \gamma \{ \delta \} }(x)=
R^{\delta_1 \alpha_1}_{\beta_1 \, \alpha}(x)
R^{\delta_2 \alpha_2}_{\beta_2 \alpha_1}(x)
\ldots R^{\delta_L \, \gamma}_{\beta_L \,\, \alpha_{L-1}}(x)=
\ba{c}
\unitlength=0.50mm
\begin{picture}(99.,29.)
\put(64.,15.){\vector(1,0){24.}}
\put(55.,15.){\makebox(0,0)[cc]{$\cdots$}}
\put(32.,-5.){\makebox(0,0)[cc]{$\s \beta_1$}}
\put(42.,-5.){\makebox(0,0)[cc]{$\s \beta_2$}}
\put(77.,-5.){\makebox(0,0)[cc]{$\s \beta_L$}}
\put(32.,5.){\makebox(0,0)[cc]{$\s 1$}}
\put(43.,5.){\makebox(0,0)[cc]{$\s 1$}}
\put(78.,5.){\makebox(0,0)[cc]{$\s 1$}}
\put(19.,19.){\makebox(0,0)[cc]{$\s x$}}
\put(32.,32.){\makebox(0,0)[cc]{$\s \delta_1$}}
\put(42.,32.){\makebox(0,0)[cc]{$\s \delta_2$}}
\put(77.,32.){\makebox(0,0)[cc]{$\s \delta_L$}}
\put(47.,15.){\line(-1,0){31.}}
\put(92.,15.){\makebox(0,0)[cc]{$\s \gamma$}}
\put(12.,15.){\makebox(0,0)[rc]{$\s \alpha$}}
\put(30.,0.){\vector(0,1){28.}}
\put(40.,0.){\vector(0,1){28.}}
\put(75.,0.){\vector(0,1){28.}}
\end{picture}
\ea  ,
\end{equation}
and then take the appropriate trace in the auxiliary space. The
arbitrary constant $ \mu $ in eq.(\ref{7})
can be used to select the boundary conditions.
Choosing $ \mu = 1 $, one obtains the $ spl_q(2,1)$-invariant
supersymmetric t-J model with open boundary conditions (OBC),
already discussed in refs. \cite{For2} \cite{Gr1}.
Other quantum group invariant integrable models, such as the $XXZ$
model \cite{Nep1}, \cite{Des}, the $SU_q(N)$ \cite{Gr2} and the
$SU_q(n/m)$ \cite{MR}, \cite{Rui} chains were also considered in
connection with OBC.

In what follows, we consider the limit  $\mu\to\infty$.
In this limit the contributions from
$ \tilde{T} $ to the monodromy matrix $ \cal{U} $
and consequently to the transfer matrix ( see eq.(\ref{11}) )
reduce to a product of constant $R$-matrices ($R_+$) ( see
eq.(\ref{5}) ). We will prove
that this choice yields a quantum group invariant supersymmetric
t-J model with PBC.

The "doubled" monodromy matrix $ \cal{U}$~(\ref{7}) can be
represented as a $ 3 \times 3 $ matrix whose entries are matrices
acting on the "quantum space"
\begin{equation}
\label{9}
{\cal{U}}^{\gamma }_{\alpha }(x) =
\left( \matrix{{\cal{A}}(x) &
\matrix{{\cal{B}}_2(x)&{\cal{B}}_3(x)}\cr        
\matrix{{\cal{C}}_2(x)\cr {\cal{C}}_3(x)}&\overline{\vrule        
\matrix{{\cal{D}}^{2}_2(x)&{\cal{D}}^2_3(x)\cr
     {\cal{D}}^3_2(x)&{\cal{D}}^3_3(x)}}}\right) \, \, \, .
\end{equation}
Using eqs.~(\ref{3}) and~(\ref{4}) ( already in the limit
$\mu\to\infty$) we can prove that it fulfills
the following Yang-Baxter relation
\begin{equation}
\label{10}
R^{\alpha \phantom{0} \beta
\phantom{0}}_{\alpha^\prime \beta^\prime}(y/x) {\cal
U}^{\beta^\prime}_{\gamma^\prime}(x)
{R_+}^{\gamma^\prime \alpha^\prime}_{\gamma \phantom{0}
\delta^\prime \phantom{0}} {\cal
U}^{\delta^\prime}_{\delta}(y)=
{\cal U}^{\alpha}_{\alpha^\prime}(y)
{R_+}^{\alpha^\prime \phantom{0} \beta}_{\delta^\prime
\beta^\prime} {\cal
U}^{\beta^\prime}_{\gamma^\prime}(x)
R^{\gamma^\prime \delta^\prime}_{\gamma \phantom{0} \delta
\phantom{0}}(y/x) \, \, \, .
\end{equation}
We observe in the
equation above the presence of constant $R$-matrices ($R_+$)
instead of spectral parameter dependent $R$-matrices, which appear
in the corresponding relation using OBC \cite{For2}, \cite{Gr1}.
This will simplify considerably the algebraic nested Bethe ansatz.

Finally, the transfer matrix is defined as the Markov trace
associated with the superalgebra ${spl}_q(2,1)$ ( $ K^\alpha_\alpha
$ ) of the " doubled" monodromy matrix
in the auxiliary space
\begin{equation}
\label{11}
{\cal T}^{ \{ \delta \} }_{ \{\beta\} }(x) =
   \sum_{\alpha} K^\alpha_\alpha
   {\cal{U}}^{\alpha \{ \delta \} }_{\alpha \{ \beta \} } \, \, \,
=    \, \, \,
\ba{c}
\unitlength=0.50mm
\begin{picture}(95.,49.)
\put(46.,15.){\makebox(0,0)[cc]{$\cdots$}}
\put(21.,-3.){\makebox(0,0)[cc]{$\s \beta_1$}}
\put(31.,-3.){\makebox(0,0)[cc]{$\s \beta_2$}}
\put(71.,-3.){\makebox(0,0)[cc]{$\s \beta_L$}}
\put(21.,47.){\makebox(0,0)[cc]{$\s \delta_1$}}
\put(31.,47.){\makebox(0,0)[cc]{$\s \delta_2$}}
\put(71.,47.){\makebox(0,0)[cc]{$\s \delta_L$}}
\put(15.,25.){\oval(20.,20.)[l]}
\put(70.,0.){\vector(0,1){45.}}
\put(30.,0.){\vector(0,1){45.}}
\put(20.,0.){\vector(0,1){45.}}
\put(73.,35.){\vector(1,0){8.}}
\put(57.,15.){\line(1,0){25.}}
\put(47.,35.){\makebox(0,0)[cc]{$\cdots$}}
\put(7.,34.){\makebox(0,0)[rc]{$\s \mu/x$}}
\put(4.,17.){\makebox(0,0)[rc]{$\s x$}}
\put(15.,15.){\line(1,0){21.}}
\put(5.00,25.){\circle{1.5}}
\put(80.,40.){\oval(10.,10.)[rb]}
\put(80.,10.){\oval(10.,10.)[rt]}
\put(92.50,40.){\oval(15.,20.)[t]}
\put(92.50,10.){\oval(15.,20.)[b]}
\put(100.,25.){\circle{1.5}}
\put(100.,41.){\line(0,-1){12.}}
\put(100.,9.){\line(0,1){22.}}
\put(30.,15.){\vector(-1,0){16.}}
\put(22.,35.){\line(1,0){6.}}
\put(32.,35.){\line(1,0){6.}}
\put(57.,35.){\line(1,0){8.}}
\put(80.,35.){\line(0,0){0.}}
\put(14.,35.){\line(1,0){4.}}
\end{picture}
\ea\ ,
\end{equation}
where
\begin{equation}
\label{12}
K^\alpha_\alpha = \sigma_\alpha \, q^{(- 2 \sum_{\gamma}^{\alpha -
1}    \sigma_\gamma) - \sigma_\alpha + 1}  \, \, \, ,
\end{equation}
and
\begin{equation}
\label{13}
\sigma = \pmatrix{1&0&0\cr
         0&1&0\cr
         0&0&-1\cr} \, \, \, .
\end{equation}
The Yang-Baxter equation for the "doubled" monodromy matrix ${\cal
U}$~(\ref{10}) implies that the transfer matrix~(\ref{11}) commutes
for different spectral parameters, which proves the integrability
of the model. Then, from the above defined transfer matrix and the
following properties
\begin{equation}
\label{14}
\ba{c}
\quad R^{\, \, \, \, \alpha^{\prime \prime} \beta^{\prime
\prime}}_{\pm \alpha^{\prime} \beta^{\prime}}
R^{\, \, \, \, \beta^\prime \alpha^\prime}_{\mp \beta \alpha} \, =
\, \delta^{\alpha^{\prime \prime}}_{\alpha}
\delta^{\beta^{\prime \prime}}_{\beta} \, \, \, \, \, \, \, \, \,
\, \, \, \, \, \, \, \, \, \, \, \, \, \, \, \, \, \, \, \, \,
\unitlength=0.45
mm
\begin{picture}(260.,40.)
\put(2.,20.){\line(1,1){8.}}
\put(32.,20.){\makebox(0,0)[rc]{$=$}}
\put(2.,0.){\line(1,1){20.}}
\put(2.,20.){\line(1,-1){8.}}
\put(14.,32.){\line(1,1){8.}}
\put(2.,40.){\line(1,-1){20.}}
\put(22.,0.){\line(-1,1){8.}}
\put(2.,40.){\vector(-1,1){1.}}
\put(22.,40.){\vector(1,1){1.}}
\put(2.,45.){\makebox(0,0)[cc]{$\s \alpha^{\prime \prime}$}}
\put(26.,45.){\makebox(0,0)[cc]{$\s \beta^{\prime \prime}$}}
\put(2.,-4.){\makebox(0,0)[cc]{$\s \alpha$}}
\put(24.,-4.){\makebox(0,0)[cc]{$\s \beta$}}
\put(72.,20.){\makebox(0,0)[rc]{$=$}}
\put(42.,20.){\line(1,1){20.}}
\put(62.,0.){\line(-1,1){20.}}
\put(82.,0.){\line(0,1){40.}}
\put(92.,0.){\line(0,1){40.}}
\put(42.,0.){\line(1,1){8.}}
\put(42.,40.){\line(1,-1){8.}}
\put(54.,12.){\line(1,1){8.}}
\put(62.,20.){\line(-1,1){8.}}
\put(62.,40.){\vector(1,1){1.}}
\put(42.,40.){\vector(-1,1){1.}}
\put(82.,40.){\vector(0,1){1.}}
\put(92.,40.){\vector(0,1){1.}}
\put(42.,45.){\makebox(0,0)[cc]{$\s \alpha^{\prime \prime}$}}
\put(66.,45.){\makebox(0,0)[cc]{$\s \beta^{\prime \prime}$}}
\put(42.,-4.){\makebox(0,0)[cc]{$\s \alpha$}}
\put(64.,-4.){\makebox(0,0)[cc]{$\s \beta$}}
\put(84.,45.){\makebox(0,0)[cc]{$\s \alpha^{\prime \prime}$}}
\put(96.,45.){\makebox(0,0)[cc]{$\s \beta^{\prime \prime}$}}
\put(81.,-4.){\makebox(0,0)[cc]{$\s \alpha$}}
\put(93.,-4.){\makebox(0,0)[cc]{$\s \beta$}} \, \, \, ,
\end{picture}
\ea\ ,
\end{equation}

\begin{equation}
\label{15}
\ba{c}
\quad R^{\, \, \, \, \, \alpha \alpha^\prime}_{\pm \alpha^\prime
\beta} K^{\alpha^\prime}_{\alpha^\prime} \, = \, q^{\pm 1} \,
\delta^\alpha_\beta \, \, \, \, \, \, \, \, \, \, \, \, \, \, \, \,
\, \, \, \, \, \, \, \, \, \, \, \, \, \, \, \, \, \, \,
\unitlength=0.3mm
\begin{picture}(110.,30.)
\put(30.,0.){\vector(-1,1){30.}}
\put(18.,18.){\vector(1,1){12.}}
\put(0.,-8.){\makebox(0,0)[cc]{$\beta$}}
\put(106.,38.){\makebox(0,0)[cc]{$\alpha$}}
\put(106.,-8.){\makebox(0,0)[cc]{$\beta$}}
\put(0.,38.){\makebox(0,0)[cc]{$\alpha$}}
\put(0.,0.){\line(1,1){12.}}
\put(30.,15.){\oval(30.,30.)[r]}
\put(78.,15.){\makebox(0,0)[cc]{$=q$}}
\put(105.,0.){\vector(0,1){30.}}
\end{picture}
\ea  ,
\quad
\ba{c}
\unitlength=0.3mm
\begin{picture}(110.,30.)
\put(0.,0.){\vector(1,1){30.}}
\put(12.,18.){\vector(-1,1){12.}}
\put(0.,-8.){\makebox(0,0)[cc]{$\beta$}}
\put(109.,38.){\makebox(0,0)[cc]{$\alpha$}}
\put(109.,-8.){\makebox(0,0)[cc]{$\beta$}}
\put(0.,38.){\makebox(0,0)[cc]{$\alpha$}}
\put(30.,0.){\line(-1,1){12.}}
\put(30.,15.){\oval(30.,30.)[r]}
\put(80.,15.){\makebox(0,0)[cc]{$=q^{-1}$}}
\put(110.,0.){\vector(0,1){30.}}
\end{picture}
\ea\ ,
\end{equation}
we obtain a quantum group invariant one-dimensional
supersymmetric t-J model with PBC through
\begin{equation}
\label{16}
{\cal H} \, \propto \, \frac{\partial}{\partial x}\ln
({\cal T})\left|_{x=1}\right. \, \, \, .
\end{equation}
This yields
\begin{equation}
\label{17}
{\cal{H}} \, = \, \sum_{j = 1}^{L-1} h_j + h_0 \, \, \, ,
\end{equation}
where
\begin{eqnarray}
 \label{18}
 h_j &=& - \sum_\sigma (
 c^{\dagger}_{j,\sigma} c_{j+1,\sigma} + c^{\dagger}_{j+1,\sigma}
c_{j,\sigma} ) - \cos \gamma n_j + 2 \cos \gamma \\
 \nonumber
 &-& 2 \biggl[ S_j^x
 S_{j+1}^x + S_j^y S_{j+1}^y + \cos \gamma \phantom{0}(S_j^z
S_{j+1}^z - {n_j n_{j+1}\over 4}) \biggr]  \\
\nonumber
 &+& i\sin( \gamma ) (n_j - n_{j+1})
 -i\sin ( \gamma)(n_j S_{j+1}^z - S_j^z
 n_{j+1}) \, \, \, ,
\end{eqnarray}
and $h_0$ is a boundary term given by
\begin{equation}
\label{19}
h_0 = \underbrace{ {\hat{R}}^{-}_{1} {\hat{R}}^{-}_{2}
\dots {\hat{R}}^{-}_{L-1}}_{\mbox{{G}} }
\, h_{L-1} \, \, \,
\underbrace{ {\hat{R}}^{+}_{L-1} \dots {\hat{R}}^{+}_{2}
{\hat{R}}^{+}_{1}}_{\mbox{{$G^{-1}$}} } \, \, \, ,
\end{equation}
with
\begin{equation}
\label{20}
{\hat{R}}^{ \pm \, \{ \gamma \} }_{ j \,\, \{ \beta \} } \, = \,
{\bf{1}}^{\gamma_1}_{\beta_1} \otimes
{\bf{1}}^{\gamma_{2}}_{\beta_{2}} \otimes \dots
{R}^{\pm \, \gamma_j \gamma_{j+1} }_{\, \, \, \, \, \, \beta_{j+1}
\beta_j} \otimes
\dots  {\bf{1}}^{\gamma_L}_{\beta_L} \, \, \, \, \, \, \, \, \, \,
\, \, \, \, j = 1, 2,  \dots L-1 \, \, \, .
\end{equation}
The presence of this boundary term $(h_0)$ is essential
to construct a quantum group invariant model with PBC.
Notice that it emerges naturally from the present construction.
The other possibility to obtain a quantum group invariant
Hamiltonian ( $h_0 = 0 $, which corresponds to OBC), was
already discussed in refs. \cite{For2}, \cite{Gr1}.
In eq. (\ref{18})  $L$ is the number of sites of the quantum chain,
$c_{j\pm}^{(\dagger)}$'s are spin up or down annihilation
(creation) operators, the $\vec S_j$'s
spin matrices and the $n_j$'s occupation numbers of electrons at
lattice site $j$.
The operators $H$, $h_i$ and ${\hat{R}}^{\pm}_i$ ($i= 1, 2,  \dots
L-1$) act on the
"quantum space" ${\bf C}^{3L}$ ( for simplicity, we omit the
quantum space indices and write them only whenever necessary).

It was shown in \cite{Zap} using methods of topological
quantum field theory that the transfer matrix obtained by this
approach for an $U_q(sl(n))$ invariant chain is equivalent to the
partition function of a vertex model
on a torus and therefrom the periodicity of that model is evident.
However, here it is not
obvious that the Hamiltonian~(\ref{17}) describes a model with PBC.
To prove this fact we first notice that the operators $
\hat{R}^{\pm}$'s are a representation of the Hecke algebra
\cite{Deg}\footnote{To obtain relations (\ref{21}), the Yang-Baxter
algebra~(\ref{3})
and eq.~(\ref{20}) have been used}
\begin{eqnarray}
\label{21}
&&{\hat{R}}^{\pm}_{j} {\hat{R}}^{\pm}_{j} = \pm (q-1/q)
{\hat{R}}^{\pm}_{j} + {\bf 1}
\, \, \, ,
\nonumber \\
&&{\hat{R}}^{\pm}_{j} {\hat{R}}^{\pm}_{j \pm 1} {\hat{R}}^{\pm}_{j}
= {\hat{R}}^{\pm}_{j \pm 1} {\hat{R}}^{\pm}_{j} {\hat{R}}^{\pm}_{j
\pm 1} \, \, \, ,
\\
\nonumber
&&{\hat{R}}^{\pm}_{i} {\hat{R}}^{\pm}_{j} =
{\hat{R}}^{\pm}_{j} {\hat{R}}^{\pm}_{i} \, \, \, \, \, \, \, \, \,
\, \, \, \, \, \, \, \, \, \, \, \,| i - j | \geq 2
\, \, \, .
\end{eqnarray}
From the Hecke algebra conditions~(\ref{21}) and the following
relation \begin{equation}
\label{22}
h_j = - \hat{R}^{\pm}_j \, \, + \, \, q^{\pm 1} {\bf 1} \, \, \, \,
\, \, \, \, \, \, \, \, \, \, j = 1, 2, \dots L-1 \, \, \, ,
\end{equation}
we find that the operator $G^{-1}$ maps $h_{j}$ into $h_{j-1}$
\begin{equation}
\label{23}
G^{-1} h_{j} G \, = \, h_{j-1 } \, \, \,
\, \, \, \, \, \, \, \, \, \, \,
j = 2, \dots L-1 \, \, \, ,
\end{equation}
and $h_1$ into $h_0$
\begin{equation}
\label{24}
G^{-1} h_{1} G \, = \, G h_{L-1} G^{-1} \, \, \, .
\end{equation}
Then, denoting by ${\cal H}_{1, 2, \dots L}$ the Hamiltonian of
eq.(\ref{17}) and by ${\cal H}_{L, 1, 2, \dots L-1}$ that one
obtained by cyclic permutation $ (1, 2, \dots L)
\rightarrow (L, 1, 2,  \dots L-1) $ and
using the properties~(\ref{23}),(\ref{24}), we show that
\begin{equation}
\label{25}
{\cal{H}}_{L,1,2, \dots L-1} \, = \,
G^{-1} {\cal{H}}_{1, 2, \dots L} G \, \, \, ,
\end{equation}
i. e., both Hamiltonians are physically equivalent, which finishes
the proof that we are dealing with a periodic chain.

Notice that, although the boundary term  (\ref{19}) is apparently
non-local, it is local in the sense that it commutes with the
local observables, in particular, the generators of the
Hecke algebra \cite{Kar1}
\begin{equation}
\label{255}
[ h_0 \, , \, {\hat{R}}^{\pm}_{j} ] \, = \, 0 \, \, \, \, \, \, \, \, \,
\, \, \, \, \, \, \, \, \, \, \, \, 1 < j  <  L-1
\, \, \, .
\end{equation}
This can be verified by using eqs. (\ref{21}) and (\ref{22}).
Finally, the
quantum group invariance of the Hamiltonian (\ref{17}) follows
directly from the fact that the operators $ \hat{R}^{\pm}$'s
are a representation of the Hecke algebra.

Next we solve the eigenvalue problem of the transfer matrix,
\begin{equation}
\label{26}
{\cal T} \Psi = ( {\cal{A}} \, + q^{-2} {\cal{D}}^2_2 \, - q^{-2}
{\cal{D}}^3_3 ) \Psi = \Lambda \Psi \, \, \, ,
\end{equation}
(and consequently of the Hamiltonian~(\ref{17})) through the
algebraic nested Bethe ansatz (ANBA) with two levels.
According to the first level Bethe ansatz, the vector $\Psi$ can be
written as
\begin{equation}
\label{27}
\Psi =\sum^{3}_{\{\alpha\}=2} {\cal{B}}_{\alpha_1}(x_1)
{\cal{B}}_{\alpha_2}(x_2)\cdots {\cal{B}}_{\alpha_r}(x_{r})
\Psi_{(1)}^{\{ \alpha \} } \Phi \, \, \, .
\end{equation}
The coefficients $\Psi_{(1)}$ are determined later by the
second level Bethe ansatz while  $ \Phi$ is the reference state
defined by the equation
$$ {\cal{C}}_{\alpha} \Phi = 0 \ \ \ \mbox{for}\ \ \
\alpha = 2,3 \ ,
$$
whose solution is $\Phi =\otimes^{L}_{i=1}{|1>}_i $. It is an
eigenstate of $\cal{A}$  and ${\cal{D}}^\alpha_\beta$
\begin{eqnarray}
\label{28}
&&{\cal{A}}(x)\Phi =q^L \, a(1/x)^{L} \Phi \, \, \, , \\
\label{29}
&&{\cal{D}}^{\alpha}_{\beta}(x)\Phi =\delta^{\alpha}_{\beta} \,
b(1/x)^L \Phi \, \, \, .
\end{eqnarray}
Following the general strategy of
the algebraic nested Bethe ansatz we apply the transfer
matrix~(\ref{11}) to the eigenvector $\Psi$~(\ref{27}). Using the
following
commutation rules, derived from the Yang-Baxter
relations~(\ref{10})
\begin{eqnarray}
\label{30}
{\cal A}(x){\cal B}_{\alpha}(y) =
{ 1 \over q } { a(x/y) \over b(x/y)}
{\cal B}_{\alpha}(y) {\cal A}(x) -
{1 \over q} { c_- (x/y) \over b(x/y)}
{\cal B}_{\alpha}(x){\cal A}(y) -
{ {q-1/q} \over q } \sum_{\beta =2}^{3}
{\cal B}_{\beta}(x) {\cal D}^{\beta}_{\alpha}(y) \, , \\
\label{31}
{\cal D}^{\gamma}_{\beta}(x) {\cal B}_{\alpha}(y) =
{R_+}^{\alpha^{\prime} \gamma}_{\gamma^{\prime} \delta^{\prime} }
{ R^{ \beta^{\prime} \gamma^{\prime}}_{\beta \, \, \alpha} (y/x)
\over b(y/x) }
{\cal B}_{\alpha^\prime}(y) {\cal
D}^{\delta^{\prime}}_{\beta^{\prime}}(x) - {R_+}^{\alpha^{\prime}
\gamma}_{\beta \beta^{\prime} }
{ c_-(y/x) \over b(y/x) }
{\cal B}_{\alpha^\prime}(x) {\cal D}^{\beta^{\prime}}_{\alpha}(x)
\, ,
\end{eqnarray}
we commute $ \cal{A}$ and $ \cal{D} $ with all $ \cal{B} $'s and
apply them to the reference state $\Phi$.
All indices in eqs.(\ref{30}) and~(\ref{31})
assume only the values $2, 3$. We begin by considering the
action of $\cal{A}$ on $\Psi$. Using eq.(\ref{30}) two types of
terms arise
when $\cal{A}$ passes through ${\cal{B}}_{\alpha}$. In the first
one, $\cal{A}$ and
${\cal{B}}_{\alpha}$ preserve their arguments and in the second,
their arguments are exchanged. The first kind of terms are called
``wanted terms'', since they can originate a vector proportional to
$\Psi$; this can not happen to the second type and therefore they
are called the "unwanted terms (u.t.)". Notice that in the present
formulation ( $\mu\to\infty$ ) the decomposition into wanted
and unwanted terms appears naturally, as in the usual periodic
case (where the transfer matrix, which is not quantum group invariant,
is constructed by taking the trace of the standard row-to-row monodromy
matrix). This is in contrast to the
case of OBC ( $\mu = 1$ ), where it is necessary to redefine the
${\cal D}$-operators in order to recognize wanted and unwanted
contributions \cite{For2}, \cite{Gr1}. After successive
applications of
eq.(\ref{30}) together with~(\ref{28}) we obtain
\begin{equation}
\label{32}
{\cal A}(x) \Psi = q^{L-r} {a(1/x)}^L \prod_{i=1}^{r}
{a(x/x_i) \over b(x/x_i)} \Psi \, + \, \mbox{u. t.} \, \, \, .
\end{equation}
Correspondingly we obtain from the commutation relations between
${\cal D}$ and ${\cal B}$~(\ref{31}) and eq.(\ref{29})
wanted and unwanted contributions
\begin{equation}
\label{33}
q^{-2} ({\cal{D}}^2_2 \, - {\cal{D}}^3_3 )
\Psi =
{b(1/x)}^L  \prod_{i=1}^{r} {1 \over b(x_i/x)}
\sum_{ \{ \alpha^{\prime} \} = 2 }^{3}
{\cal{B}}_{\alpha^\prime_1}(x_1)
{\cal{B}}_{\alpha^\prime_2}(x_2)\cdots
{\cal{B}}_{\alpha^\prime_r}(x_{r}) q^{-1} {\cal T}^{
\{\alpha^\prime \} }_{(1)}
\Psi_{(1)} \, + \, \mbox{u. t.} \, \, \, .
\end{equation}
Here we have introduced a new (second level) transfer matrix
\begin{equation}
\label{34}
{{\cal T}}_{(1)}=\sum^{3}_{\alpha=2} \sigma_{\alpha} \, q^{-1} \,
{{\cal{U}}_{(1)}}^{\alpha}_{\alpha} \, \, \, ,
\end{equation}
as the Markov trace associated to the superalgebra
$ SU_q(1,1)$ of the second level "doubled" monodromy matrix ${\cal
U}_{(1)}$, defined analogously to ${\cal U}$ ( see eq.~(\ref{9}) ).
Now, all indices
range from $2$ to $3$, as in the
internal block of the matrix ${ \cal U}$~(\ref{9}).
Thus, we will treat the internal block ${\cal D}$ in the same way
as we have done with the whole matrix, through the
identification
${\cal A}_{(1)} \equiv {{\cal U}_{(1)}}_{2}^{2},\, \,
{\cal B}_{(1)} \equiv {{\cal U}_{(1)}}_{3}^{2}, \, \,
{\cal C}_{(1)} \equiv {{\cal U}_{(1)}}_{2}^{3}$ and
${\cal D}_{(1)} \equiv {{\cal U}_{(1)}}_{3}^{3} $,
The first term ( wanted term ) on the
right-hand side of eq.(\ref{33}) is proportional to $\Psi$ if the
eigenvalue equation
\begin{equation}
\label{35}
{\cal T}_{(1)}\Psi_{(1)}=\Lambda_{(1)}\Psi_{(1)}
\end{equation}
is fulfilled. The eigenvector $\Psi_{(1)}$ of ${\cal T}_{(1)}$ is
defined by the second level Bethe Ansatz
\begin{equation}
\label{36}
\Psi_{(1)} = {\cal{B}}_{(1)}(y_1)
{\cal{B}}_{(1)}(y_2) \cdots {\cal{B}}_{(1)}(y_m)
\Phi_{(1)} \, \, \, ,
\end{equation}
where $\Phi_{(1)}$ is the second level reference state
given by $\Phi_{(1)}=\otimes^{r}_{i=1}|2>_{i}$, as a result of
being annihilated by ${\cal {C}}_{(1)}$. Then, proceeding along the
same lines as in the previous step, we apply
${\cal T}_{(1)}$~(\ref{34}) to the state $\Psi_{(1)}$~(\ref{36})
and pass ${\cal A}_{(1)}$ and ${\cal D}_{(1)}$ through the
${{\cal B}_{(1)}}^\prime $ s, using commutation relations
derived from the Yang-Baxter relation~(\ref{10}) and the action of
${\cal A}_{(1)}$ and ${\cal D}_{(1)}$ on the vacuum $\Phi_{(1)}$.
As before, we obtain wanted and unwanted contributions
\begin{eqnarray}
\label{37}
{\cal A}_{(1)}(x) \Psi_{(1)} = q^{-m + r}
\prod_{i=1}^{r}a(x_i/x)
\prod_{j=1}^{m} {a(x/y_j) \over b(x/y_j) } \Psi_{(1)} \, +
\, \mbox{u. t.} \, \, \, , \\
\label{38}
{\cal D}_{(1)}(x) \Psi_{(1)} = (-1)^m q^{-m}
\prod_{i=1}^{r}b(x_i/x)
\prod_{j=1}^{m} {w(y_j/x) \over b(y_j/x) } \Psi_{(1)} \, +
\, \mbox{u. t.} \, \, \,  .
\end{eqnarray}
Then, combining eqs.(\ref{37}),(\ref{38}),(\ref{33}),(\ref{32})
and~(\ref{26}) we get the eigenvalue $\Lambda(x)$ of the transfer
matrix ${\cal T}$  if the " unwanted terms " cancel out
\begin{eqnarray}
\label{39}
\Lambda(x) &=& q^{L-r} {a(1/x)}^L \prod_{i=1}^{r} { a(x/x_i) \over
b(x/x_i) } + q^{-2 + r -m} {b(1/x)}^L \prod_{i=1}^r { a(x_i/x)
\over b(x_i/x)} \prod_{j=1}^m { a(x/y_j) \over b(x/y_j)} \\
\nonumber
&-& (-1)^m q^{-2-m}{b(1/x)}^L \prod_{j=1}^{m} { w(y_j/x) \over
b(y_j/x)} \, \, \, . \end{eqnarray}
All unwanted terms vanish if the Bethe ansatz equations
hold. They can be obtained by demanding that the eigenvalue
$\Lambda(x)$~(\ref{39}) has no poles at $x = x_i$ and
$x = y_j $, since ${\cal T}$ is an analytical function in $x$
\begin{eqnarray}
\label{40}
q^{L+m+2-2r} {\biggl( {a(1/x_k) \over b(1/x_k) }  \biggr)}^L
\prod_{i=1}^{r} { a(x_k/x_i) \over b(x_k/x_i) }
{b(x_i/x_k) \over a(x_i/x_k) }
\prod_{j=1}^{m} {b(x_k/y_j) \over a(x_k/y_j)} = -1 \, \, \, , \, \,
\, \, \, \, \, \, \, \, \, \, \, \,
k = 1, \dots r \, \, \, , \\
\label{41}
(-1)^m q^r \prod_{i=1}^{r} {a(x_i/y_l) \over b(x_i/y_l)}
\prod_{j=1}^{m} {a(y_l/y_j) \over b(y_l/y_j)} {b(y_j/y_l) \over
w(y_j/y_l)} = 1 \, \, \, , \, \, \, \, \, \, \, \, \, \, \, \, \,
\,
l = 1, \dots m \, \, \, .
\end{eqnarray}
Therefore, we have reduced the eigenvalue problem of the transfer
matrix ${\cal T}$ to a system of coupled algebraic equations for
the parameters $x$ and $y$. Notice that these equations are much
simpler than those obtained for OBC ( see refs. \cite{For2},
\cite{Gr1} ).
A possible application of these results ( eqs. (\ref{40}), (\ref{41}) )
would be an analysis of the structure of the ground state and
some low lying excitations of the model in the thermodynamic limit.
The question of the completeness of the Bethe states for an $spl_q(2,1)$
invariant supersymmetric t-J model is left open. In ref. \cite{For1},
this point was treated for the isotropic case ( $q=1$ ), where
a complete set of eigenvectors was obtained by combining the
Bethe ansatz with the $spl(2,1)$ underlying supersymmetry of
the model. The completeness of its $q$-deformed version
is under investigation.

\section*{Acknowledgements}

The author would like to thank M. Karowski for useful discussions
and CNPq ( Conselho Nacional de Desenvolvimento Cient\'{\i}fico e
Tecnol\'ogico ) for financial support.

\newpage

\end{document}